\documentclass[trackchanges, twocolumn]{aastex701}
\usepackage{amsmath} 
\usepackage{siunitx}

\newcommand{\micromu}{\si{\micro Hz}}

\begin{document}

\title{Numerical Simulations Confirm Wave-Induced Shear Mixing in Stellar Interiors}

\author[sname='North America']{A. Varghese}
\affiliation{Universit\'e Paris-Saclay, Universit\'e  Paris Cit\'e, CEA, CNRS, AIM, F-91191, Gif-sur-Yvette, France}
\email[show]{ashlin.varghese@cea.fr}  

\author[gname=Savannah,sname=Africa]{R.P. Ratnasingam}
\affiliation{Newcastle University, Newcastle Upon Tyne, United Kingdom}
\email{rathish.ratnasingam@newcastle.ac.uk}

\author[gname=Savannah,sname=Africa]{L. Ramírez-Galeano}
\affiliation{University of Geneva, Geneva, Switzerland}

\email{laura.ramirezgaleano@unige.ch}

\author[sname='North America']{S. Mathis} 
\affiliation{Universit\'e Paris-Saclay, Universit\'e  Paris Cit\'e, CEA, CNRS, AIM, F-91191, Gif-sur-Yvette, France}
\email{stephane.mathis@cea.fr}

\author{T.M. Rogers}
\affiliation{Newcastle University, Newcastle Upon Tyne, United Kingdom}
\affiliation{Planetary Science Institute, Tucson, AZ, USA}
\email{tamirogers@mac.com}

\begin{abstract}

Internal Gravity Waves (IGWs) are 
thought to cause mixing in stellar interiors, a process that has been widely studied both theoretically and numerically. Our aim is to determine the physical mechanism responsible for the wave-induced mixing in stellar interiors. We compare  the mixing profiles obtained from two-dimensional (2D) equatorial hydrodynamical and tracer particle simulations with theoretical predictions from \cite{gs1991} and \cite{zahn_92} on wave mixing  due to wave-induced shear turbulence. 
Our results show that, { despite not satisfying the vertical shear instability threshold}, the mixing profiles from the simulations agree remarkably well with the theoretical predictions of {both prescriptions}, 
strongly suggesting that shear from IGWs plays an important role in mixing even at low shear rates. 
This agreement remains robust across different stellar masses, ages, rotation and simulation parameters. This {provides an important step in providing realistic parameterisations for wave mixing in stellar structure and  evolution models.} 

\end{abstract}



\section{Introduction}

Mixing processes in stellar radiative interiors affects the distribution of chemicals,  
influencing the lifetime, evolution and the final stages of stars \citep{maeder_physics_2009}. Constraining  this internal mixing is necessary for the accurate modelling of stellar structure and evolution. This has been an ongoing challenge for decades and various mixing mechanisms { have been} proposed as the source of additional mixing in stars to improve  stellar evolution models \citep{mathi_2013, CRS}. Our understanding of stellar interiors has  improved significantly with the recent advances in asteroseismology, which have provided new  insights into properties such as internal rotation and mixing from the core to the surface \citep{aerts_asteroseismic_2024}. 
Among the various internal processes proposed to cause mixing, Internal Gravity Waves (IGWs)  are a leading candidate. These waves were already known to play a major role in the transport of energy and momentum in the Earth's atmosphere and oceans \citep{holton, alex_2010}.

 In stars, IGWs are { thought} to transport angular momentum \citep{scht_1993, zan97,  kumar} and chemicals \citep{Press, gs1991, TC5}.  This process has been widely explored theoretically for the last $\sim$ 40 years. \cite{gs1991} { proposed that the shear induced by IGW could produce the weak turbulence needed to }
cause mixing, that could possibly explain the Li depletion in the Sun and F-type stars. Their model assumes a Kolmogorov spectrum for the turbulent eddies at the convective-radiative boundary  and an incoherent distribution of convective eddies exciting monochromatic waves that propagate into the stable layer.  They argued that in a steady flow, the vertical shear induced by IGWs can lead to small-scale turbulence capable of driving mixing, with a IGW diffusion coefficient that scales  with the square of the wave amplitude, determined by the Richardson number and thermal diffusion. \cite{zahn_92} estimated the diffusion coefficient for shear-induced mixing in stably stratified flows, based on a detailed analysis of anisotropic turbulence focusing on large scale shears. Although not specific to wave-driven flows, \cite{zahn_92}'s prescription similarly assumes that the largest eddies dominate vertical transport and retrieve a diffusion coefficient similar to that of \cite{gs1991}. 

Apart from turbulence, \cite{Press} proposed that the waves can also induce mixing  through transport effects related to the non-conservation of entropy. He considered a turbulent flow described by the largest eddies of sizes at the scale of mixing length which excites monochromatic oscillation in the radiation zone.  In the presence of thermal diffusion, the fluid elements undergo irreversible oscillations due to the loss of entropy and fail to  return to their original position, resulting in a mean squared displacement. By relating this displacement to the wave-induced velocity, he derived a diffusion coefficient with a fourth order dependence on wave amplitude, for wave mixing associated with the resulting Stokes displacement. This contrasts with the second order dependence proposed for the wave-induced shear mixing by \cite{gs1991}. 
Following the work of \cite{Press}, later studies improved this approach by incorporating a turbulent spectrum with Kolmogorov scaling, considering a superposition of monochromatic waves excited at the convective radiative interface and modeling the wave excitation by convective plumes rather than by convective eddies that follow the mixing length theory (MLT). These models have been used to explain the Li depletion in solar-type stars \citep{scht_1993, mont94, MS96, mons_2000}.  
Taking into account these two distinct mechanisms of wave-induced mixing, \cite{denissenkov_partial_2003} demonstrated that wave mixing by either of these mechanisms could be an additional source of mixing in AGB stars. 
 
 Building on these theoretical works, multidimensional hydrodynamic simulations have  explored the excitation and propagation of waves in both solar type and massive stars \citep{rogers_gravity_2005, rogers_internal_2013, alvan_2014, custon, edelmann_three-dimensional_2019, ratnasingam_two-dimensional_2020, breton22, herwig_3d_2023, morton}. Numerical studies 
 {have further investigated} the transport of angular momentum by these waves 
 \citep{RG2006, barker_ogl, rogers_internal_2013, rogers2025} and have  successfully explained the asteroseismically inferred internal rotation rates 
 \citep{rogers2015, aerts2021}. The wave amplitude spectra predictions from the numerical simulations \citep{rogers_internal_2013, edelmann_three-dimensional_2019} were also found to be consistent with the observed brightness variation in O and B type stars \citep{apj2015,bownman2019}.

Such simulations have also been used to investigate chemical mixing by IGWs  following different approaches for obtaining the  diffusion coefficient. Based on the 3D hydrodynamic simulation of a  25M$_{\odot}$, \cite{herwig_3d_2023} determines a diffusion coefficient in 1D by solving an inverted diffusion equation using a radial profile of the composition as the input \citep{jones2017}. \cite{Rogers2017} conducted 2D hydrodynamical simulations of an intermediate mass star with a convective core and radiative envelope and introduced tracer particles in their simulations considering an equatorial geometry. By tracking the trajectories of these particles, they determined the mean squared displacement and the diffusion coefficient for wave mixing. Following their work, \cite{varghese_2023} extended the analysis across age and mass.  These simulations shows the variation of mixing profile across the radius in the stellar interior and  also provided a theoretical prescription based on their findings across different masses and ages. \cite{joey_ar} studied the evolution of the Nitrogen surface abundance ratio \citep{hunter, martin2024} by implementing this mixing prescription in the one dimensional stellar evolution code MESA \citep{mesa1, mesa2, mesa3, mesa4, mesa5}. 
 
 The next important step is to gain  insight into  the dominant physical mechanism responsible for the wave mixing.  While \cite{Rogers2017} found empirically that the wave mixing was proportional to the wave amplitude squared, they did not link that to any previous theory. \cite{herwig_3d_2023}  
 attempted to  determine if the mixing caused by IGWs arises from  the wave-induced shear instabilities \citep{gs1991, zahn_92} and found that the measured mixing by IGWs did not agree with the shear mixing predictions. 
 More recently, \cite{morton} investigated the different wave mixing mechanisms proposed by \cite{gs1991} and \cite{Press} by conducting fully compressible 2D hydrodynamical simulations considering  20M$_{\sun}$ main sequence star assuming a meriodional geometry. 
 They too found that their simulation results did not match either theory, though the \cite{gs1991} theory had a very similar radial profile.  Notably, both the simulations of \cite{herwig_3d_2023} and \cite{morton} use an implicit large eddy simulation (ILES) approach and while they explicitly set their thermal diffusion coefficients, it is very likely that those values are not accurate and that the thermal diffusion of the simulation is numerical, similar to their viscosity.  This is discussed at length in \cite{herwig_3d_2023}, therefore estimating the actual thermal diffusion to use in making these direct comparisons is somewhat fraught. Moreover, the comparison made by \cite{morton} only included one frequency when comparing to and examined theories for diffusion, propagation and generation, making it difficult to isolate which theory was not in agreement with simulations.\footnote[1]{ We note that the ``clumping" problem for particle tracking seen in \cite{morton} does {\it not} occur in our simulations.  We expect that this behavior is due to small-scale discontinuities due to their numerical method {(see Appendix \ref{clumping})}.}

In this work we aim to compare the theoretical prescriptions for wave-induced shear mixing proposed by \cite{gs1991} and \cite{zahn_92} 
 with the { diffusion coefficients retrieved in \cite{Rogers2017} and \cite{varghese_2023} using tracer particles combined with 2D hydrodynamical equatorial simulations.  We explicitly isolate the theory for diffusion. } 
 Section \ref{theory} will discuss the theoretical prescriptions proposed by  \cite{gs1991} and \cite{zahn_92}. Section \ref{methods} will discuss the numerical simulation profiles used in this study. Comparison of these  mixing profiles from the numerical studies with the theoretical prescriptions  are presented in Section \ref{compare} and a detailed discussion on the findings and conclusions are presented in Section \ref{discn}.

\section{Theoretical prescriptions} \label{theory}

\subsection{Mixing by wave-induced Shear \citep{gs1991, zahn_92}.}
 A steady flow with sinusoidally varying vertical shear produces small scale turbulence \citep{zahn75}. In a stably stratified region, vertical mixing occurs when the shear is strong enough to overcome the stabilizing effect of the buoyancy in the vertical direction. A flow is said to be unstable when the Richardson number, $Ri$, that compares the effect of the buoyancy force with that of the strength of the shear is less than $1/4$; the Richardson number is defined as: 
 \begin{equation}\label{ri_con}
    R_i = \frac{N^2}{\left({\partial u_h}/{\partial z}\right)^2} 
\end{equation}
\begin{figure*}
    \centering
    \includegraphics[width=1\linewidth]{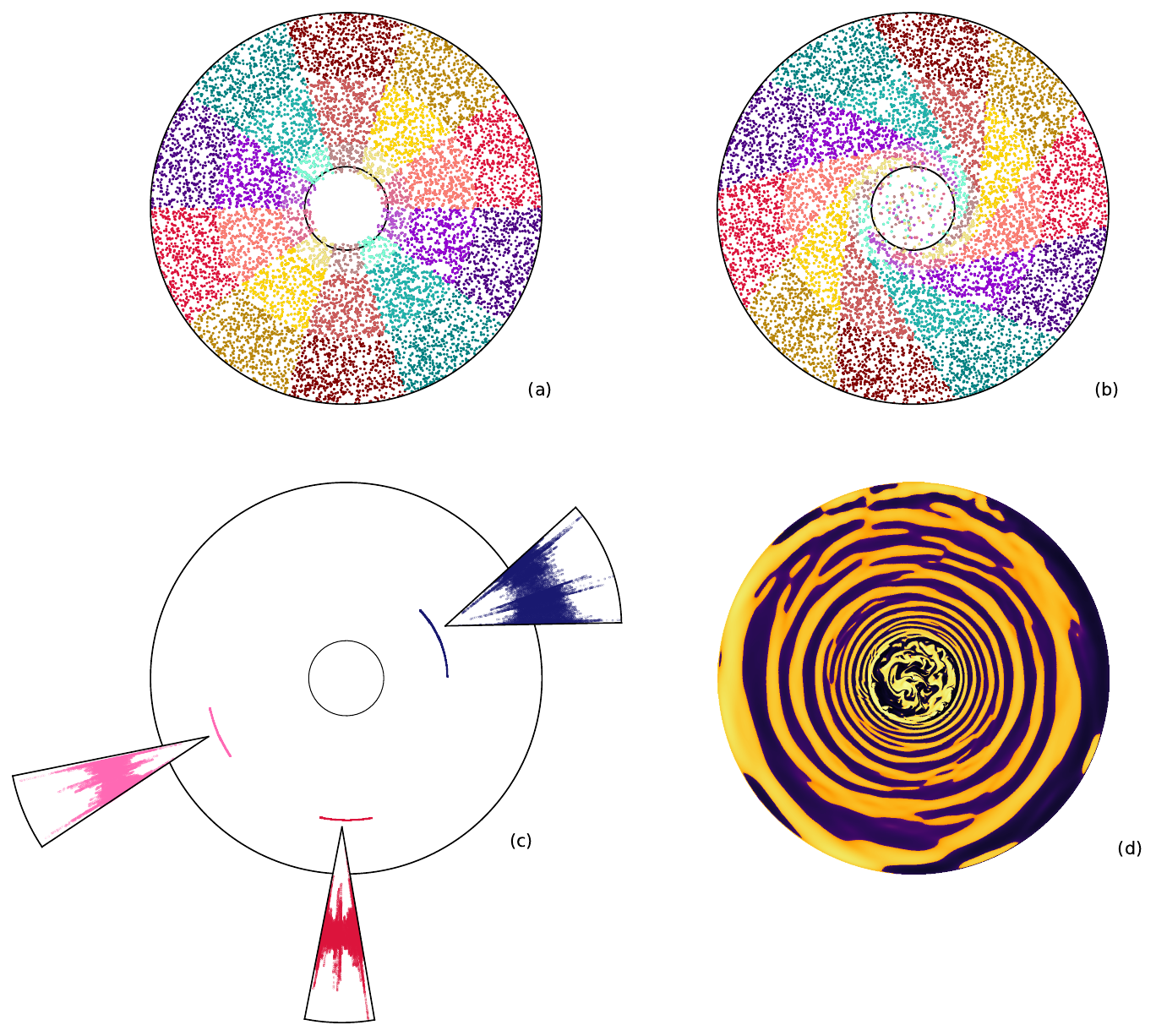}
    \caption{(a) Initial and (b) final particle distribution from the tracer particle simulations. The color scheme is chosen to visualize the particle motion with time. (c) The trajectory of a single particle over the whole simulation domain at 0.5R$_{\star}$ (blue), 0.72R$_{\star}$ (red)and 0.7R$_{\star}$ (pink). (d) Vorticity from the 2D simulations with positive vorticity (yellow) saturated at $1 \times$ 10$^{-4}$ s$^{-1}$ and negative vorticity (blue) saturated at $-1 \times$ 10$^{-4}$ s$^{-1}$. This is for the 7M$_{\odot}$ ZAMS model from \cite{varghese_2023}. }
    \label{plt_vort}
\end{figure*}

 \noindent where $N$ is the Brunt-V\"{a}is\"{a}l\"{a} frequency, $u_h$ is the horizontal velocity and $z$ is the vertical distance. However, the above is valid only in the absence of thermal diffusion. Thermal diffusion reduces the stabilizing effect of stratification by weakening the restoring force. \cite{town} pointed out that when the thermal timescale, $\tau_t = l^2/K$, of the turbulent element is shorter than its turnover time, $\tau_{ov} = l/v$, where $l$ is the size of the fluid element, $v$ is the velocity of the fluid element and $K$ is the thermal diffusivity, the criterion for instability is modified to:
 \begin{equation}\label{mod_ri}
       \frac{N^2}{\left({\partial u_h}/{\partial z}\right)^2} \frac{\tau_{t}}{\tau_{ov}}\leq \frac{1}{4}.
 \end{equation}
 Building on this understanding of how thermal diffusion influences the shear stability, \cite{zahn_92} determined the  vertical diffusion coefficient due to turbulent motion. In the presence of a strong horizontal turbulence, 
 the transport of chemicals in the vertical direction triggered by the vertical shear instabilities and meridional flows behaves as a diffusive process \citep{chab_zah}. The turbulent motions are stronger in the horizontal direction than in the vertical direction since the latter is inhibited by the buoyant forces \citep{zahn1983}. Assuming that the vertical transport is dominated by the largest turbulent eddies that satisfy the modified Richardson criterion for shear instability, 
 \cite{zahn_92} proposed the diffusion coefficient associated with turbulence generated by vertical shear as
\begin{equation}\label{dzhan}
    D_{zahn} = \mathcal{C}\left(\frac{du_h}{dr}\right)^2\frac{K}{N^2},
\end{equation}
where the exact value of $\mathcal{C}$ is unknown and  the thermal diffusivity, $K$, is defined as
\begin{equation}\label{ther_dif_gls}
    K = \frac{16}{3}\frac{\sigma T^3}{\kappa \rho^2 c_p}.
\end{equation} 
Assuming that the wave mixing is due to the wave-induced vertical shear and the parameter $\mathcal{C} \sim 0.1$ as determined from the hydrodynamic simulations of \cite{prat_2016}, we compute the local wave diffusion coefficient based on the above equation in section \ref{compare}.
 
\cite{gs1991} noted that the horizontal oscillation of gravity waves creates a sinusoidally varying vertical shear which  can lead to  weak turbulence. Accounting for the energy losses by waves due to the work done by the weak turbulence against stable stratification, along with the radiative damping and assuming that the mixing occurs on scales where Eqn.~\eqref{mod_ri} is satisfied, \cite{gs1991} proposed a diffusion coefficient\footnote[2]{{Note the equivalence of Eqn. \ref{gls} with Eqn. \ref{dzhan} in the case of low-frequency IGWs for which $\vert {\rm d}/{\rm d}r\vert \sim k_v$ \citep[see ][]{Mathis2005}}.} for the wave-induced mixing,
\begin{equation}\label{gls}
    D_{GLS} = \frac{K}{4N^2}(k_v u_h)^2,
\end{equation}

where $k_v$ is the vertical wavenumber. 

{\cite{gs1991} considered the following inequality
\begin{equation}\label{crit_gls}
    k_vu_h > \omega
\end{equation}
as a sufficient requirement for their proposed weak mixing mechanism, noting that this criterion must be satisfied for the shear flow produced by IGWs of frequency $\omega$ to be treated as approximately steady state\footnote[3]{{Note that Eqn. \ref{crit_gls}  is equivalent to the breaking criterion proposed by \cite{Press} and derived by \cite{mathis2025}}.}. 
However, \cite{gs1991} explicitly note that mixing at lower shear rates was conceivable. 

We employ the above two approaches to estimate the diffusion coefficient. The prescription by \cite{zahn_92} uses the horizontal velocity directly from the simulation. In contrast, \cite{gs1991} focus on the wave contribution by decomposing the velocity into frequency and wavenumber components using a Fourier transform. While the underlying form of the equation is the same in both cases, the latter approach specifically targets the wave-driven component of the flow and implicitly includes a time average over the simulation time.

\section{Mixing profiles from 2D hydrodynamics and Tracer Particle simulations}\label{methods}
For the comparison of the theoretical prescription with the simulations, we use the { diffusion} 
profiles presented in \cite{varghese_2023} for 3, 7 and 20 M$_{\odot}$ at ZAMS and midMS. 
The simulations use the background reference state models from MESA and were run for a total simulation time of $\sim$ 3$\times$10$^{7}$s considering constant thermal ($\kappa$) and viscous ($\nu$) diffusivities (see Table \ref{table_1} in Appendix \ref{clumping}). 
These profiles are compared with the theoretical prescriptions discussed in section \ref{theory}. Figure \ref{plt_vort} (a) and (b) shows the initial and final particle distributions from the tracer particle simulation. The trajectory traversed by the particle over the whole simulation time at three different radius is shown in (c) and  the vorticity\footnote[4]{Vorticity is defined as $\nabla \times \vec{v}$, where $\vec{v}$ is the fluid velocity.} from the 2D simulations is shown in (d). The figure shows a 7M$_{\odot}$ ZAMS model representing the 
same simulations from \cite{varghese_2023}.

\section{Comparison of the theory with that of the simulation} \label{compare}
\subsection{Wave-Induced shear mixing}
We initially calculated the local diffusion coefficient based on \cite{zahn_92}, using Eqn.~\eqref{dzhan} at a given time and along different horizontal directions. This is shown in Fig.~\ref{plt_zn_gls_3z} (blue) for a particular horizontal direction. On comparison with that of the simulation profiles (black), we find that these theoretical profiles follow a similar trend as that of the simulation diffusion profiles. {We note that the periodic decrease in amplitude occurs when the shear of the wave, i.e., the radial derivative of the horizontal velocity goes to zero.} 
Although this is at one time snapshot and horizontal direction, we found that this agreement is similar for all times and horizontal locations.  

We then calculated the diffusion coefficient proposed by \cite{gs1991} based on the wave-induced vertical shear turbulence, using the Eqn.~\eqref{gls} for frequencies in the range of 3 -- 13 \micromu{} and wavenumber, $l$ = 1. To obtain the wave velocities, we decomposed the horizontal velocity from the simulation into its frequency and wavenumber components using a Fourier transform in space and time. The range of frequencies and wavenumber were chosen based on  the dominant frequencies determined by \cite{varghese_2023}. 
We determined the contribution of individual frequencies to the diffusion coefficient and { summed over frequencies to retrieve the ``GLS" diffusion coefficient, {D$_{GLS-sum}$,
\begin{equation}\label{gls_sum}
    D_{GLS-sum}(r) = \frac{K}{4N^2}\sum_{\nu=3}^{13 \micromu{}}\left(k_v u_h(r,\omega,m)\right)^2,
\end{equation}
shown in red in Figure \ref{plt_zn_gls_3z}}}. We found that this profile  gives us a reasonable agreement between the theory {\footnote[5]{{In the GLS prescription, the periodic decrease of the diffusion coefficient occurs when the horizontal velocity itself changes sign.}}} and the simulations . {We chose this wavenumber and frequency range as we found these to provide the dominant contribution to mixing in \cite{varghese_2023}. Moreover,} 
summing over a broader range of frequencies and wavenumbers did not significantly alter the  amplitude of the profile. In both cases, we see that the theories from \cite{zahn_92} and \cite{gs1991} give similar results, which is expected since \cite{gs1991} represents effectively the time average of \cite{zahn_92} for low-frequency progressive IGWs. {The theoretical diffusion coefficient calculated using Eqn.~\eqref{gls_sum}, but with the thermal diffusivities determined by MESA are shown in orange in Fig.~\ref{plt_zn_gls_3z}.  This will be discussed further in section \ref{obsvn}. } The green dashed lines in Fig.~\ref{plt_zn_gls_3z} is based on the study of \cite{cope2020} and will be discussed in detail in Section \ref{discn}.

\begin{figure*}
\includegraphics[width=0.96\textwidth]{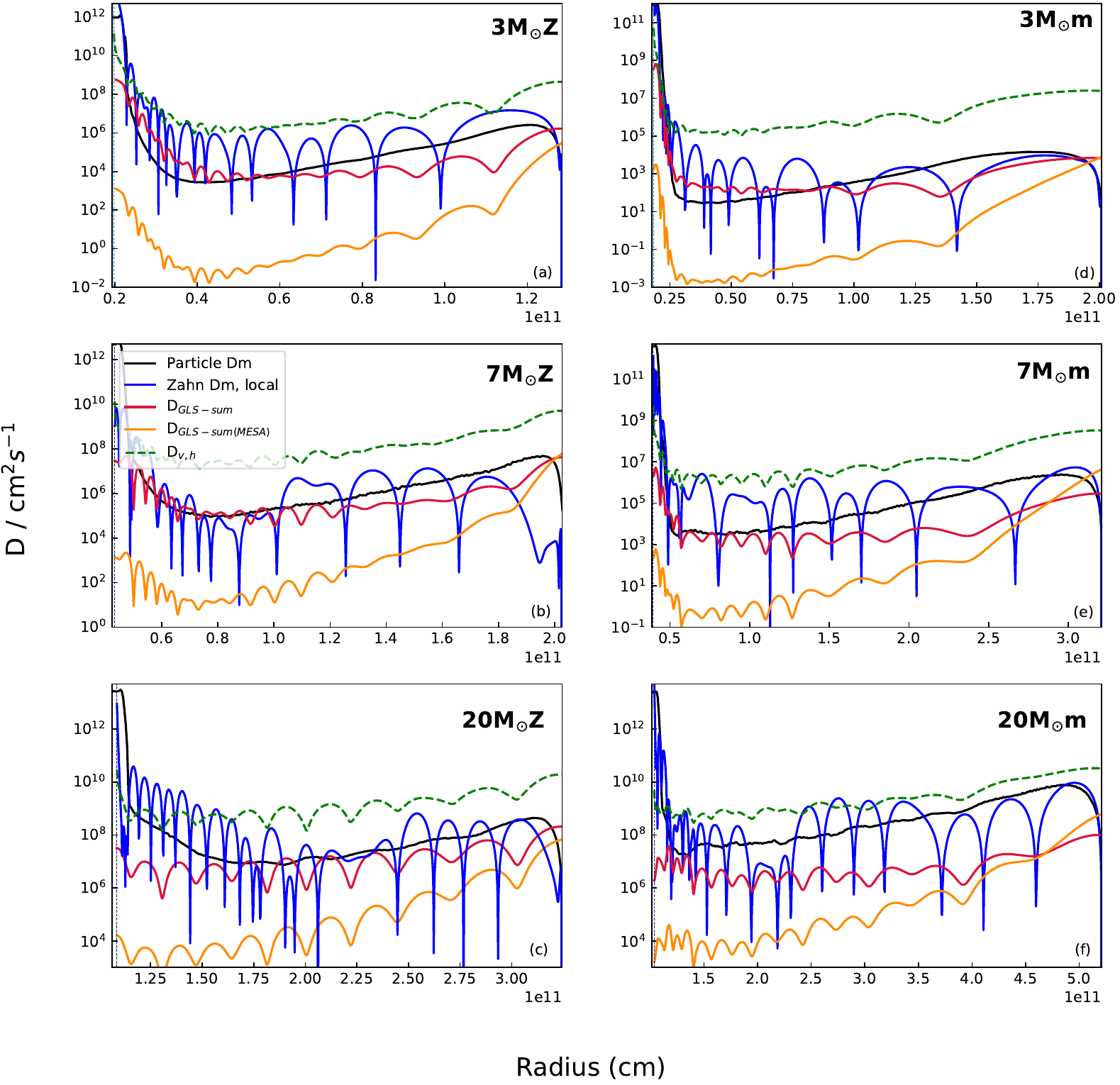}
\caption{Diffusion coefficient from the tracer particle simulation (black) along with the theoretical profiles calculated using  Eqn.~\eqref{dzhan} based on \cite{zahn_92} (blue) and Eqn.~\eqref{gls} based on \cite{gs1991} (red) using the horizontal velocity (u$_h$) from the simulation as a function of  radius for 3M$_{\odot}$ (top), 7M$_{\odot}$(middle) and 20M$_{\odot}$ (bottom)  for the ZAMS models (left, represented with `Z') and midMS models (right, represented with `m'). {The diffusion coefficeint calculated from Eqn.~\eqref{gls} but considering realisitic thermal diffusivities from MESA is shown in orange}. The green dashed line (D$_{v,h}$) is based on Eqn.~\eqref{dv_hor}.}
\label{plt_zn_gls_3z}
\end{figure*}
\subsection{Significance for Observations}\label{obsvn}
Gaining an understanding of the physical mechanism driving the  wave mixing is a major step towards the improvement of stellar evolution models. Recently, \cite{joey_ar} implemented the wave mixing in the 1D stellar evolution model, MESA. They followed the parameterization proposed by \cite{Rogers2017} for the diffusion coefficient as  
\begin{equation}\label{rog_d}
    D=Av_{wave}^2,
\end{equation}
where `A' is an unknown parameter speculated to  depend on various factors, including thermal diffusivity.  Their study calibrated this parameter and found that `A'  must {increase} with stellar mass in order to reproduce the observed nitrogen abundances. 

The agreement between theory and simulation presented in this work provides insight into the value of the parameter `A'. Comparing Eqn.~\eqref{rog_d} with Eqn.~\eqref{gls}, `A' could be written as ,
\begin{equation}\label{aval}
    A \sim \frac{Kk_v^2 }{4N^2} 
\end{equation}
which is dependent on the thermal diffusivity.   Fig.~\ref{Aplot} shows the radial profile of `A' calculated using this relation but with realistic thermal diffusivity values from MESA instead of the constant (higher) values used in simulations. The frequencies chosen correspond to the dominant frequencies determined in \cite{varghese_2023}. We find that `A' increases with stellar mass, consistent with the trend predicted by the observations \citep{joey_ar}. This allows the direct computation of `A' for implementing wave mixing in stellar evolution models, taking in to account the underlying physical mechanism.  

However, the amplitude of the diffusion coefficient obtained from the simulation also depends { on the wave velocity squared and the wave velocity depends on both the convective velocity {\it and} the thermal and viscous diffusivities in the simulations.  Accounting for the convective velocity is straight forward. Our velocities are typically 5-15 times larger than MLT and hence, {\bf if} that convective velocity is transferred directly to wave velocity, our diffusion coefficient might be 25-100 times larger than in the actual star.  However, there are numerous uncertainties in MLT velocities \citep{joycetayar23}.  Moreover, accounting for the effect of enhanced diffusivities on the wave amplitudes is non-trivial.  First,  these enhanced numerical diffusivities contribute to a larger drop in velocity across the convective-radiative interface than would be expected in an actual star.  Second, once convective perturbations transition to waves, the larger thermal and viscous diffusion will damp different frequency waves differently.  In both cases the enhanced numerical thermal and viscous dissipation damp low frequencies more, hence reducing their amplitudes and contribution to the measured diffusion coefficient near the convective boundary. On the other hand, thermal and viscous diffusions have little impact on frequencies higher than $\sim$ 10\micromu{} and hence likely do not affect the measured diffusion coefficient near the stellar surface. {This can be observed in Fig. \ref{plt_zn_gls_3z}, where the profiles using MESA  (orange) and the constant (red) thermal diffusivity have a closer agreement near the surface.} Taken together, we expect that the diffusion coefficient {obtained using the MESA thermal diffusivity (orange profile)} measured near the convective-radiative boundary is a lower limit, while that near the surface is likely close to stellar values.} Despite {the limitations with these extrapolations}, the diffusion coefficients obtained from our simulations are in the expected range inferred from  asteroseismic  {inferences of $10^2-10^6$ cm$^2$ s$^{-1}$ for near-core layers} \citep{may2018, aerts_asteroseismic_2024, brinkmann_2025}, {and around $10^5-10^6$ cm$^2$ s$^{-1}$ in the envelope based on surface abundance observations \citep{joey_ar} and detached eclipsing binaries \citep{brinkmann_2025}}. 
\begin{figure}
    \centering
    \includegraphics[width=1\linewidth]{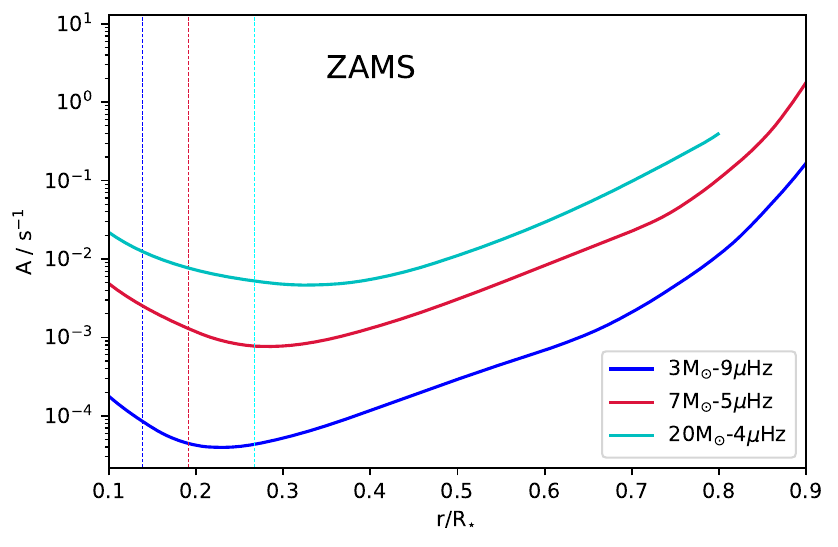}
    \caption{The parameter `A' calculated across masses using Eqn.~\eqref{aval} as a function of fractional radius.}
    \label{Aplot}
\end{figure}

\section{Discussions and Conclusions}\label{discn}

The agreement of the theoretical profiles based on \cite{gs1991}/\cite{zahn_92} with the simulation profiles presented in \cite{varghese_2023} supports the hypothesis that the underlying mechanism for wave mixing observed in the simulations is due to wave-induced vertical shear. We tested this across a range of simulations, varying mass, age, rotation rate (see Appendix \ref{rot}), convective forcing, thermal diffusion and viscosity and found similar agreement in all cases. Notably, the theoretical predictions also agree well with our recent results of solar type stars with a convective envelope and a radiative core (see Fig.~\ref{plt_zn_gls_sun}). We also highlight that the mixing coefficient near the convective-radiative boundary (at a radius $\sim$ $4.5\times 10^{10}$ cm) is $\sim$ $10^4 - 10^6$, which matches the mixing coefficient proposed to explain the Li depletion in the Sun \citep{gs1991}. The background reference state for the solar models used here are obtained from the 1D stellar evolution code STAREVOL \citep{amard2016, amard2019}. 
\begin{figure}
    \centering
    \includegraphics[width=1\linewidth]{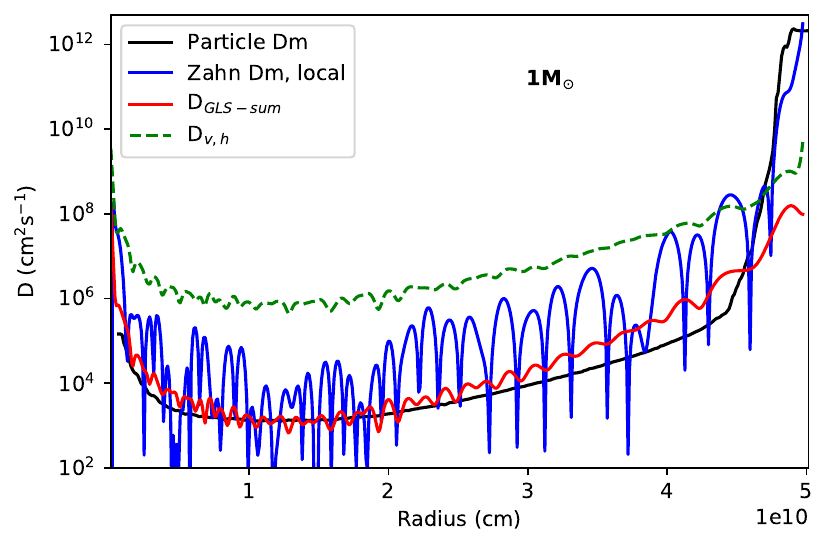}
    \caption{Diffusion coefficient from the tracer particle simulation (black) along with the theoretical profiles calculated using  Eqn.~\eqref{dzhan} based on \cite{zahn_92} (blue) and Eqn.~\eqref{gls} based on \cite{gs1991} (red) using the horizontal velocity (u$_h$) from the simulation as a function of  radius for 1M$_{\odot}$  model at solar age. The green dashed (D$_{v,h}$) line is based on Eqn.~\eqref{dv_hor}.}
    \label{plt_zn_gls_sun}
\end{figure}

Even though the theoretical prescription of wave-induced shear mixing agrees well with the simulation mixing profiles, the conditions required for developing shear instability, whether based on the criterion from \cite{gs1991} (Eqn.~\ref{crit_gls}) or the Richardson number being less than 1 (Eqn.~\ref{ri_con}) are not satisfied everywhere in our simulations. Indeed, if there were an instability, the diffusion coefficients would be significantly higher and inconsistent with observations. Therefore, we conclude that mixing is instigated by vertical shear but that {\it weak mixing} does not require instability.

We also extended our comparison to the model proposed by \cite{Press}, who derived a diffusion coefficient proportional to $u_h^4$, based on the Stokes displacement, but found that it does not match  our simulation profiles (see Fig.~\ref{scz_allz} in Appendix). We are not testing alternative diffusion due to IGWs breaking \citep{mathis2025}, since we do not observe such phenomena in our simulations.   

Turbulent transport in the vertical direction can also arise from 3D motions driven by horizontal shear instabilities \citep{zahn_92, mathis_2018}, which may contribute to the observed vertical mixing. This has been numerically tested in several studies \citep{cope2020, garaud2020, garaud_combined_2024}. \cite{cope2020} found that in thermally diffusive regions, the vertical mixing coefficient can be related to the horizontal turbulence as,
\begin{equation}\label{dv_hor}
    D_{v,h} \propto \left(\frac{u_hK}{L_h^3N^2}\right)^{1/2}u_hL_h
\end{equation}
where $L_h$ is of the order of the horizontal wavelength ($\sim$ $k_h^{-1}$). This could possibly explain the presence of vertical transport even when classical vertical shear instability conditions are not satisfied. {The above corresponds to a fully 3D configuration, where the radial, latitudinal, and azimuthal directions are all taken into account. In our case, we make use of the fact that, in the theory of linear stellar oscillations, $u_\theta$ and $u_\phi$ are of the same order of magnitude \citep{lee_saio, mathis_2009}. We therefore substitute the azimuthal velocity component from our simulation for $u_h$ in Eqn. \ref{dv_hor}, which should provide the correct order of magnitude.} The diffusion profile calculated using Eqn.~\eqref{dv_hor} is shown in Figure \ref{plt_zn_gls_3z} (green dotted lines). We observe a similar trend to that of the simulation diffusion profile (black). Although the amplitude differs more than in the  profiles obtained using \cite{zahn_92} and \cite{gs1991} prescriptions, the similar trend supports our argument that wave-driven weak turbulence  can cause mixing in stellar interiors. 

The overall consistency across different theoretical models and simulations makes weak turbulent mixing a strong candidate. Implementation of these theoretical prescriptions in 1D stellar models can bring us a step closer to calibrating  the free parameters in these models using the observational data from the current and upcoming asteroseismic space missions \citep{plato}. However, the fact that the classical Richardson number criterion is not met in these simulations suggests that an additional mechanism may be contributing to the wave-driven mixing.  Preliminary work indicates that wave-wave interactions of low-frequency propagating waves with higher frequency standing modes may contribute, but further investigation is required.

\begin{acknowledgments}
We thank the anonymous referee for their constructive comments, which allowed us to improve our work. A.V and S.M. acknowledges 
support from the European Research Council (ERC) under the Horizon Europe
programme (Synergy Grant agreement 101071505: 4D-STAR). S.M acknowledges the support from the CNES
SOHO-GOLF and PLATO grants at CEA-DAp, and from PNST (CNRS/INSU).
While partially funded by the European Union, views and opinions expressed
are however those of the author only and do not necessarily reflect those of the
European Union or the European Research Council. Neither the European Union
nor the granting authority can be held responsible for them. T.M Rogers and R.P. Ratnasingam were funded under STFC grant ST/W001020/1. The hydrodynamical simulations were carried out  on the STFC funded DiRAC Data Intensive service at Leicester (DIaL), operated by the University of Leicester IT Services, which forms part of the STFC DiRAC HPC  Facility (www.dirac.ac.uk), funded by BEIS capital funding via STFC capital grants ST/K000373/1 and ST/R002363/1 and STFC DiRAC Operations grant  ST/R001014/1 and on the Alfven facility of CEA Paris-Saclay.    
\end{acknowledgments}


\appendix

\begin{figure}
     \centering
     \includegraphics[width=0.8\linewidth]{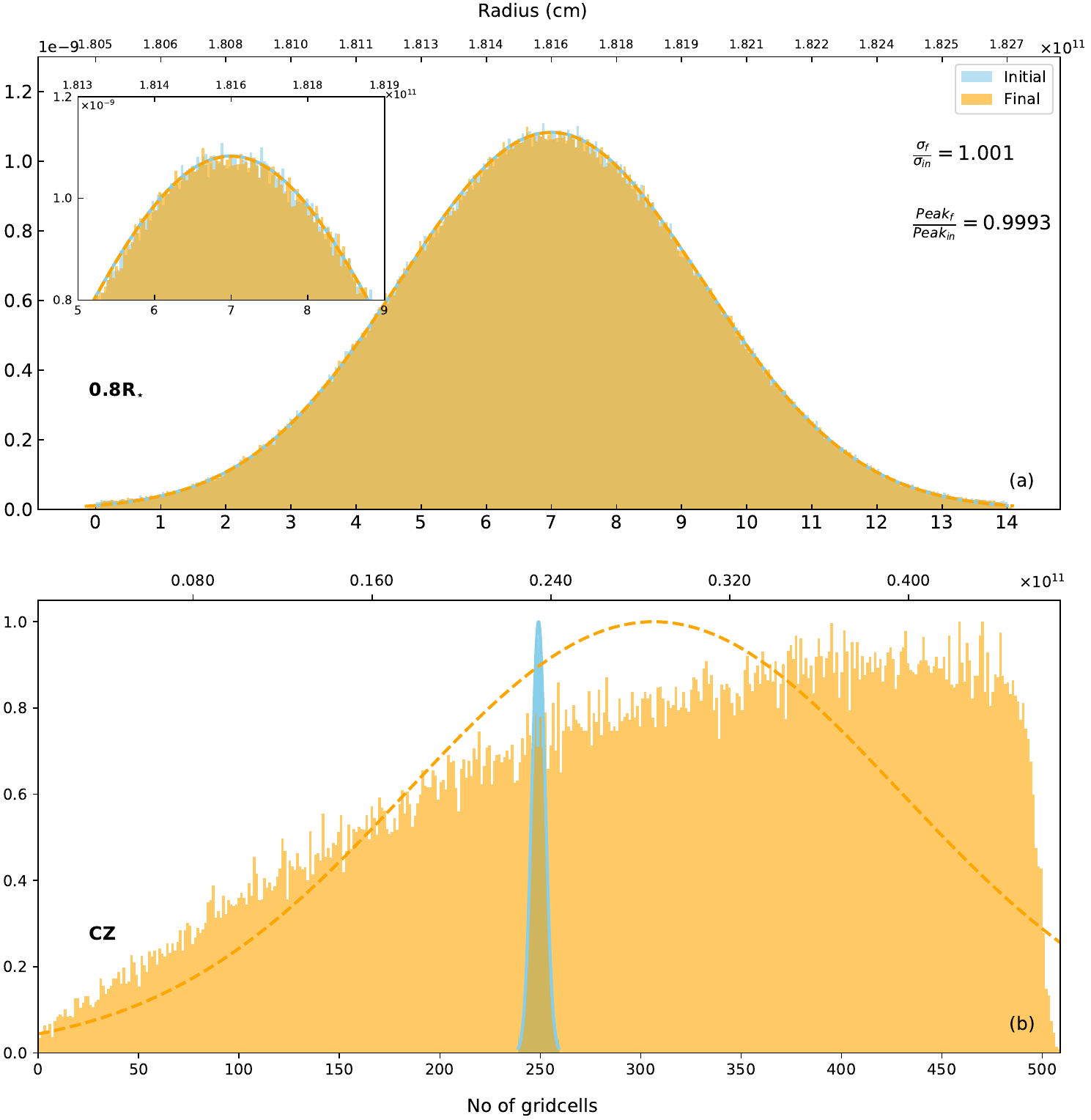}
     \caption{{Initial (blue) and final (orange) particle distributions for the 7M$_{\odot}$ ZAMS model  as a function of gridcells. (a) Distributions centered around 0.8R$_{\star}$. The ratio of final-to-initial widths is $\sigma_{f}$/$\sigma_{in}$= 1.001, and the ratio of peak heights is 0.9993. (b) Distributions within the convection zone (cz), normalized for visual comparison. For both panels, the top axis indicates the corresponding stellar radius within which particles are distributed.}}
     \label{g08}
 \end{figure}

 \begin{figure*}
     \centering
     \includegraphics[width=0.8\textwidth]{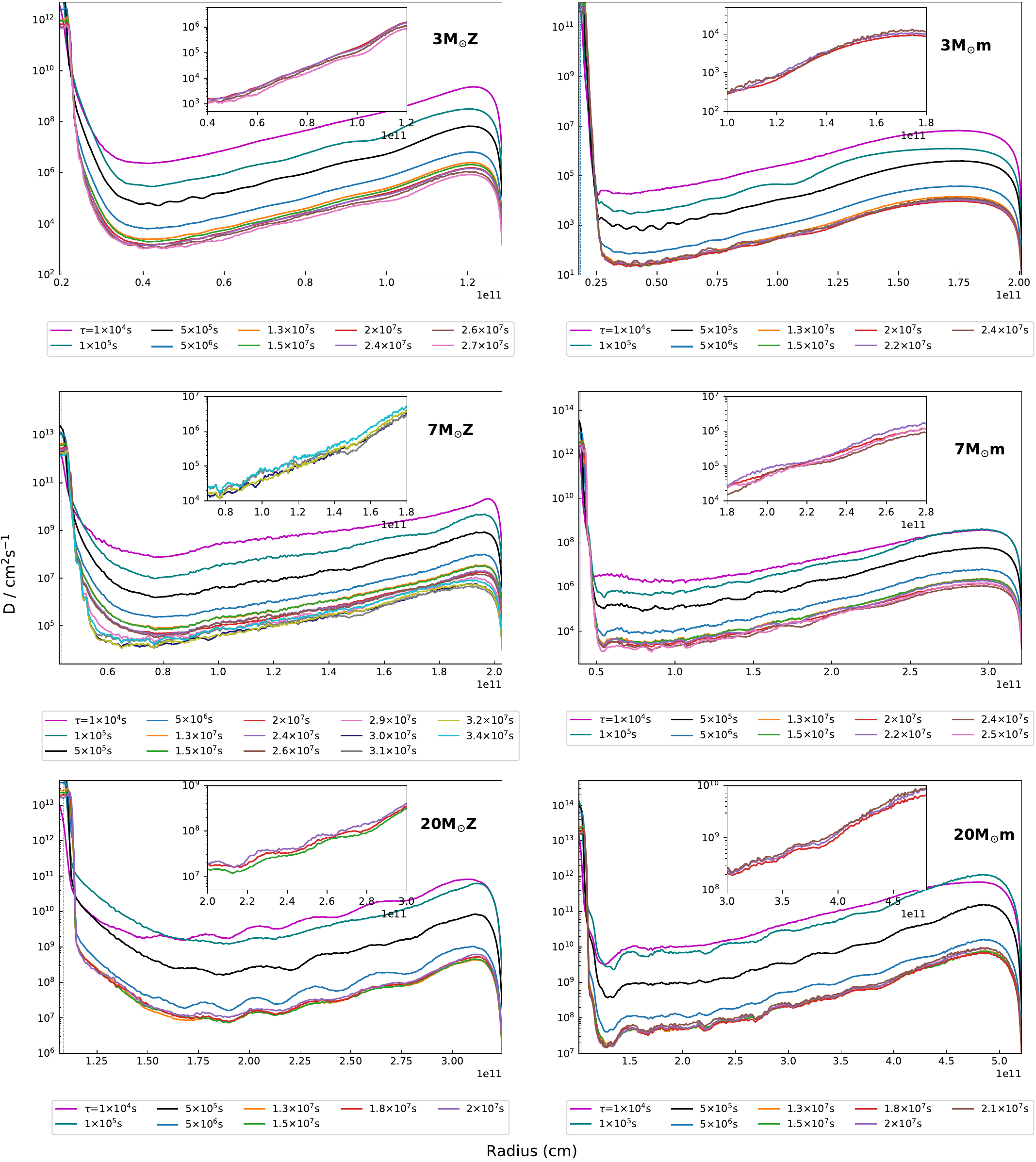}
     \caption{{The diffusion coefficient as a function of total radius at different time differences ($\tau$) for 3M$_{\odot}$ , 7M$_{\odot}$ and 20M$_{\odot}$ ZAMS (left, represented with `Z') and midMS (right, represented with `m') models. The insets highlight the zoomed-in profiles at the latest $\tau$ values. }}
     \label{dz_all}
 \end{figure*}

{\section{Particle clumping and Convergence of Mixing profiles from the Simulations}\label{clumping}}
{Recently, \cite{morton} have raised concern over using tracer particles to measure a diffusion coefficient. This concern is based on particle clumping observed in their simulations (see Figure 9 in that paper).  The authors surmise that this issue is due to the particles not following the Eulerian velocity over long time frames.  As pointed out by \cite{popov08}, \cite{genel13} and others, this mis-match between Eulerian velocities and particle (Lagrangian) velocities is most severe at discontinuities in flows (though clumping at cell centers is not typical and indicates a separate problem). To demonstrate that our simulations do not suffer from such numerical artefacts, we repeat the calculation in \cite{morton} by  initializing particles with a Gaussian distribution centered at 0.8R$_{\star}$ in our 7M$_{\odot}$ ZAMS model. Figure \ref{g08} (a) shows the initial and final distributions of 10$^6$ particles as a function of grid cell and radius. The initial Gaussian profile is preserved at the end of the simulation and no clumping is seen.  Though imperceptible on the figure, the Gaussian spreads very slightly over time, consistent with a diffusive behavior. Using the variation in the mean and peak, we calculated the diffusion coefficient directly from the spread of the Gaussian and recover the precise diffusion coefficient at that radius as seen in Figure~\ref{plt_zn_gls_3z}.  We note that a Gaussian profile is {\it not} retained for particles in the convection zone, as expected (Figure \ref{g08} (b)).}

  { To demonstrate that the diffusion profiles have converged for all models, we present the radial diffusion profiles at different time differences ($\tau$) for the ZAMS (left) and mid-MS (right) models in Figure \ref{dz_all}. The profiles remain essentially unchanged beyond $\tau \sim 2\times 10^7$ (except for 7 M$_{\odot}$ ZAMS which required longer time to converge), as seen in the zoomed-in panels for all models. The radial diffusion profile of the 7 M$_{\odot}$ ZAMS model shown here is an updated version of \cite{varghese_2023}, that was run longer in order to reach equilibrium.  Overall, the converged profiles agree well with the prescriptions of \cite{gs1991} and \cite{zahn_92}.}

{ We have reproduced the thermal and viscous diffusivities of the models from \cite{varghese_2023} in Table~\ref{table_1} for readers' convenience.
  \begin{table}[ht]
  \centering
\begin{tabular}{ccc} 
 \hline
 Model & $\kappa$, $\nu$ / $10^{12}$ $\mathrm{cm^2 s^{-1}}$  \\ [0.5ex] 
  \hline
3M$_{\odot}$ ZAMS&  5  \\
3M$_{\odot}$ midMS & 5 \\
3M$_{\odot}$ TAMS & 5  \\
7M$_{\odot}$ ZAMS & 5 \\
7M$_{\odot}$ midMS & 5  \\
7M$_{\odot}$ TAMS & 2.5 \\
20M$_{\odot}$ ZAMS & 8  \\
20M$_{\odot}$ midMS &8  \\
20M$_{\odot}$ TAMS & 5 \\ 
 \hline
\end{tabular}
\caption{{Thermal and viscous diffusivities of the models from \cite{varghese_2023}.}}
\label{table_1}
\end{table}}

{\section{Comparison of the theory with that of the simulations for models with different rotation rates}\label{rot}
Figure~\ref{plt_zn_gls_rot} shows the comparison between the diffusion profiles from the numerical simulations of \cite{varghese_effect_2024} (black) at different rotation rates ( $\Omega$ = $1\times 10^{-5}$, $2\times 10^{-5}$, $3\times 10^{-5}$, $4\times 10^{-5}$ and $1\times 10^{-4}$ rads$^{-1}$ ) and the mixing profiles computed in this work using the theoretical prescription of \cite{zahn_92} (blue) and \cite{gs1991} (red) for the 7M$_{\odot}$ midMS model.
\begin{figure*}
\includegraphics[width=0.96\textwidth]{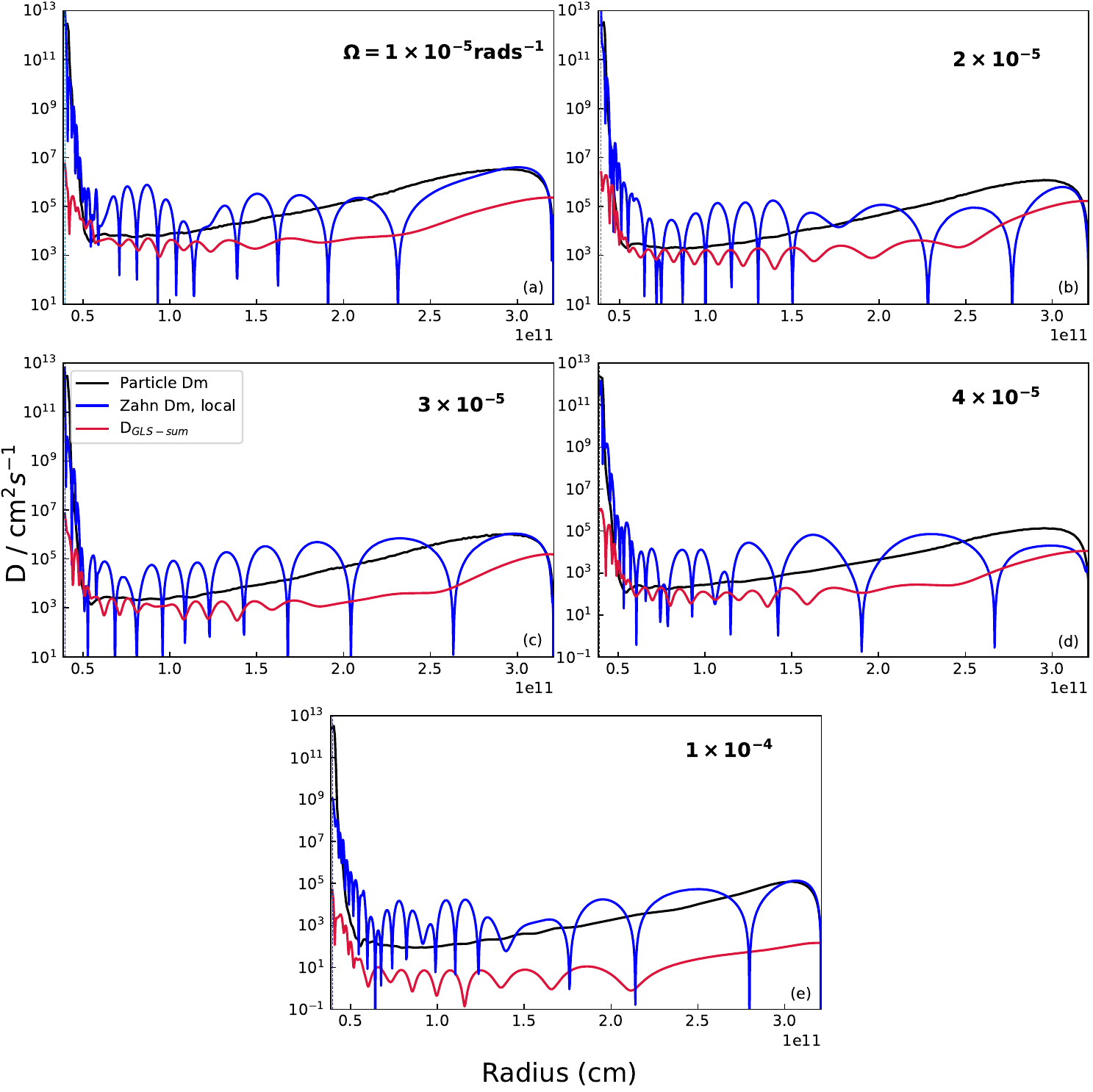}
\caption{{Diffusion coefficient from the tracer particle simulation (black) along with the theoretical profiles calculated using  Eqn.~\eqref{dzhan} based on \cite{zahn_92} (blue) and Eqn.~\eqref{gls} based on \cite{gs1991} (red) using the horizontal velocity (u$_h$) from the simulation as a function of  radius for the  7M$_{\odot}$ midMS model at rotation rates (a) $1\times 10^{-5}$ (b) $2\times 10^{-5}$ (c) $3\times 10^{-5}$ (d) $4\times 10^{-5}$ and (e) $1\times 10^{-4}$ rad$\mathrm{s}^{-1}$ ($0.13$, $0.27$, $0.41$, $0.55$ and $1.37$ d$^{-1}$ respectively).}}
\label{plt_zn_gls_rot}
\end{figure*}}
\section{Comparison of wave mixing with other theoretical prescriptions}
\subsection{\cite{Press}}
The diffusion coefficient proposed by \cite{Press} taking into account the distance traveled by the fluid element due to the loss of entropy is ,
\begin{equation}\label{dschatz}
    D_{\rm{P81}} = \frac{k_h^6u_h^4N^{2}K^2}{\omega^{7}}.
\end{equation}

\begin{figure}
    \centering
    \includegraphics[width=0.5\linewidth]{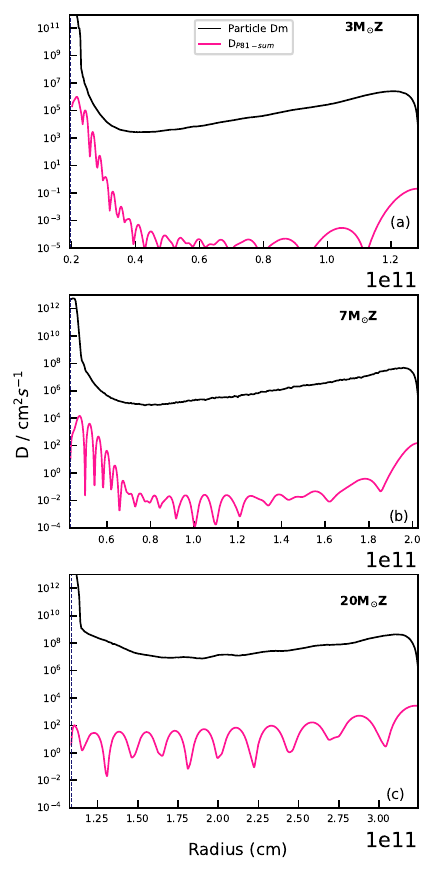}
    \caption{The diffusion coefficient from the tracer particle simulation (black) along with the theoretical profile calculated using Eqn.~\eqref{dschatz} (pink) using the horizontal velocity (u$_h$) from the simulations for (a) 3M$_{\sun}$ (b) 7M$_{\sun}$ and (c) 20M$_{\sun}$ ZAMS models as a function of radius.}
    \label{scz_allz}
\end{figure}
We calculated the theoretical diffusion coefficients using Eqn.~\eqref{dschatz} across different frequencies for the models presented in this work. Fig.~\ref{scz_allz} shows the simulation profile in black along with theoretical profile based on \cite{Press} summed over the dominant frequencies (pink). As observed, the amplitude of the profile is $\sim$ $7-8$ orders of magnitude lower than that of the simulation profiles.

\bibliography{sample701}{}
\bibliographystyle{aasjournalv7}



\end{document}